\documentclass[twoside,twocolumn,9pt]{article}

\usepackage[super,sort&compress,comma]{natbib} 
\usepackage[version=3]{mhchem}
\usepackage[left=1.5cm, right=1.5cm, top=1.785cm, bottom=2.0cm]{geometry}
\usepackage{balance}
\usepackage{mathptmx}
\usepackage[mathcal]{eucal}
\usepackage{sectsty}
\usepackage{graphicx} 
\usepackage{lastpage}
\usepackage{cuted}
\usepackage[format=plain,justification=justified,singlelinecheck=false,font={stretch=1.125,small,sf},labelfont=bf,labelsep=space]{caption}
\usepackage{float}
\usepackage{fancyhdr}
\usepackage{fnpos}
\usepackage[english]{babel}
\addto{\captionsenglish}{%
  
}
\usepackage{array}
\usepackage{droidsans}
\usepackage{charter}
\usepackage[T1]{fontenc}
\usepackage[usenames,dvipsnames]{xcolor}
\usepackage{setspace}
\usepackage[compact]{titlesec}
\usepackage{hyperref}
\usepackage[normalem]{ulem}
\usepackage[utf8]{inputenc}
\DeclareUnicodeCharacter{2009}{\,} 
\DeclareUnicodeCharacter{22EF}{\dots} 

\definecolor{cream}{RGB}{222,217,201}

\begin{document}

\pagestyle{fancy}
\thispagestyle{plain}
\fancypagestyle{plain}{
\renewcommand{\headrulewidth}{0pt}
}

\makeFNbottom
\makeatletter
\renewcommand\LARGE{\@setfontsize\LARGE{15pt}{17}}
\renewcommand\Large{\@setfontsize\Large{12pt}{14}}
\renewcommand\large{\@setfontsize\large{10pt}{12}}
\renewcommand\footnotesize{\@setfontsize\footnotesize{7pt}{10}}
\makeatother

\renewcommand{\thefootnote}{\fnsymbol{footnote}}
\renewcommand\footnoterule{\vspace*{1pt}%
\color{cream}\hrule width 3.5in height 0.4pt \color{black}\vspace*{5pt}} 
\setcounter{secnumdepth}{5}

\makeatletter 
\renewcommand\@biblabel[1]{#1}            
\renewcommand\@makefntext[1]%
{\noindent\makebox[0pt][r]{\@thefnmark\,}#1}
\makeatother 
\renewcommand{\figurename}{\small{Fig.}~}
\sectionfont{\sffamily\Large}
\subsectionfont{\normalsize}
\subsubsectionfont{\bf}
\setstretch{1.125} 
\setlength{\skip\footins}{0.8cm}
\setlength{\footnotesep}{0.25cm}
\setlength{\jot}{10pt}
\titlespacing*{\section}{0pt}{4pt}{4pt}
\titlespacing*{\subsection}{0pt}{15pt}{1pt}

\fancyfoot{}
\fancyfoot[LO,RE]{\vspace{-7.1pt}\includegraphics[height=9pt]{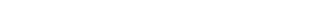}}
\fancyfoot[CO]{\vspace{-7.1pt}\hspace{11.9cm}\includegraphics{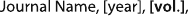}}
\fancyfoot[CE]{\vspace{-7.2pt}\hspace{-13.2cm}\includegraphics{head_foot/RF}}
\fancyfoot[RO]{\footnotesize{\sffamily{1--\pageref{LastPage} ~\textbar  \hspace{2pt}\thepage}}}
\fancyfoot[LE]{\footnotesize{\sffamily{\thepage~\textbar\hspace{4.65cm} 1--\pageref{LastPage}}}}
\fancyhead{}
\renewcommand{\headrulewidth}{0pt} 
\renewcommand{\footrulewidth}{0pt}
\setlength{\arrayrulewidth}{1pt}
\setlength{\columnsep}{6.5mm}
\setlength\bibsep{1pt}

\makeatletter 
\newlength{\figrulesep} 
\setlength{\figrulesep}{0.5\textfloatsep} 

\newcommand{\topfigrule}{\vspace*{-1pt}%
\noindent{\color{cream}\rule[-\figrulesep]{\columnwidth}{1.5pt}} }

\newcommand{\botfigrule}{\vspace*{-2pt}%
\noindent{\color{cream}\rule[\figrulesep]{\columnwidth}{1.5pt}} }

\newcommand{\dblfigrule}{\vspace*{-1pt}%
\noindent{\color{cream}\rule[-\figrulesep]{\textwidth}{1.5pt}} }

\makeatother

\twocolumn[
  \begin{@twocolumnfalse}
{\includegraphics[height=30pt]{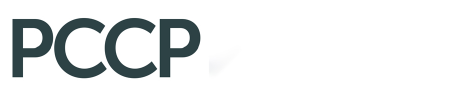}\hfill\raisebox{0pt}[0pt][0pt]{\includegraphics[height=55pt]{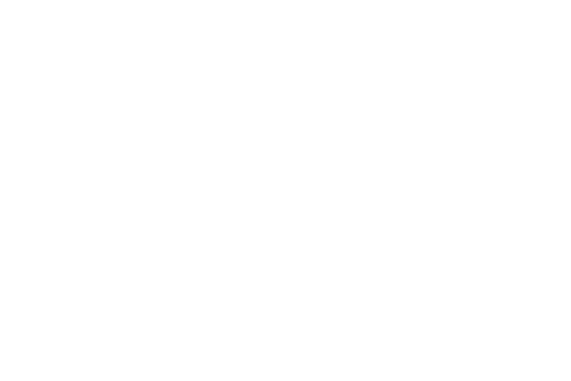}}\\[1ex]
\includegraphics[width=18.5cm]{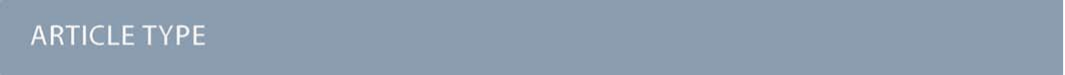}}\par
\vspace{1em}
\sffamily
\begin{tabular}{m{4.5cm} p{13.5cm} }

\includegraphics{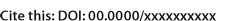} & \noindent\LARGE{\textbf{Momentum Rescaling for Collapse to a Block Ehrenfest Dynamics}} \\
\vspace{0.3cm} & \vspace{0.3cm} \\

 & \noindent\large{Ari R. X. Pereira\textit{$^{a}$} and Benjamin G. Levine$^{\ast}$\textit{$^{a}$}} \\

\includegraphics{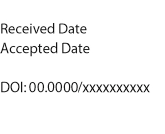} & \noindent\normalsize{ The 
   Ehrenfest with collapse to a block (TAB) algorithm has recently been demonstrated to efficiently and accurately simulate nonadiabatic molecular dynamics in dense manifolds of electronic states. TAB employs an Ehrenfest force for the classical nuclei, accompanied by stochastic collapses of the electronic density matrix to account for decoherence. Energy conservation dictates that the nuclear momentum be adjusted during such a collapse.  In this paper, we present a prescription for rescaling the component of the nuclear momentum that projects into the branching plane between arbitrary superposition states. This prescription yields accurate branching ratios and phase-space distributions when compared to alternative methods in which scaling is performed in the nonadiabatic coupling direction or the momentum direction. We justify this direction by deriving an effective nonadiabatic coupling vector from the localized Pechukas force during a collapse step.} \\

\end{tabular}

 \end{@twocolumnfalse} \vspace{0.6cm}

  ]

\renewcommand*\rmdefault{bch}\normalfont\upshape
\rmfamily
\section*{}
\vspace{-1cm}


\footnotetext{\textit{$^{a}$~Institute for Advanced Computational Science and Department of Chemistry, Stony Brook University, Stony Brook, New York 11794, United States}}
\footnotetext{$^{\ast}$~Email: ben.levine@stonybrook.edu}


\section{Introduction}
Bigger and better computers have made the direct, on-the-fly, simulation of nonadiabatic dynamics accessible\cite{levineIsomerizationConicalIntersections2007,schuurmanDynamicsConicalIntersections2018a}. A variety of methods, at varying levels of approximations, have been used to simulate ultrafast photochemistry. One approach is to use trajectory basis functions to represent the nuclear wavepacket. Methods employing this approach include the variational multiconfigurational Gaussian (vMCG) method, ab initio multiple spawning (AIMS), and cloning- based approaches. \cite{ worthGaussianWavePackets2020,ben-nunInitioMultipleSpawning2000,makhovInitioMultipleCloning2014,shalashilinQuantumMechanicsBasis2009} While these methods can accurately account for decoherence from first principles, they can be expensive to employ for systems with many populated electronic states. Another family of methods used to simulate large systems are those that employ classical trajectories for the evolution of the nuclei\cite{janosPerspectiveChallengePredicting2026}. Among these methods are surface hopping and its various versions thereof, coupled-trajectory methods, and Ehrenfest dynamics\cite{tullyMolecularDynamicsElectronic1990,tullyTrajectorySurfaceHopping1971,jainEfficientAugmentedSurface2016,agostiniQuantumclassicalNonadiabaticDynamics2016,ehrenfestBemerkungUeberAngenaeherte1927,liInitioEhrenfestDynamics2005,minCoupledtrajectoryQuantumclassicalApproach2015}.
One well known drawback of independent classical trajectory methods is that they do not naturally account for decoherence, and therefore benefit from the inclusion of decoherence corrections\cite{bittnerQuantumDecoherenceMixed1995,bittnerDecoherentHistoriesNonadiabatic1997,granucciCriticalAppraisalFewest2007,shuDecoherenceItsRole2023,zhuCoherentSwitchingDecay2004}.
Another challenge for surface hopping methods is their inability to simulate an initial coherent superposition of states, stimulating recent attempts to remedy the problem \cite{villasecoarribasNonadiabaticDynamicsClassical2024,grellModelingEvolutionLaserInduced2025}.
Furthermore, many laser experiments initially excite a molecule to superpositions of high lying electronic states\cite{kaufmanLongLivedElectronicCoherences2023}. The number of electronic states involved poses a practical problem to current nonadiabatic methods. 

With the goal of efficiently simulating systems with a large number of states while also allowing for the study of coherences, our group has proposed the collapse to a block (TAB) family of methods\cite{eschStatepairwiseDecoherenceTimes2020,eschDecoherencecorrectedEhrenfestMolecular2020,eschAccurateNonempiricalMethod2021,suchanLonglivedElectronicCoherences2025}. The TAB method uses Ehrenfest dynamics to propagate classical nuclei, accompanied by stochastic collapses of the electronic state to account for decoherence. 
Augmenting Ehrenfest dynamics with a decoherence correction is not new; prior approaches include deterministic schemes that modify the mean-field equations to damp coherences \cite{akimovCoherencePenaltyFunctional2014,zhuNonBornOppenheimerTrajectoriesSelfconsistent2004,zhuNonBornOppenheimerLiouvillevonNeumann2005,zhuCoherentSwitchingDecay2004} and stochastic schemes that collapse the electronic state at discrete events\cite{nijjarEhrenfestClassicalPath2019,prezhdoMeanFieldApproximation1999,prezhdoMeanfieldMolecularDynamics1997,bedard-hearnMeanfieldDynamicsStochastic2005,tomazEhrenfestDynamicsSpontaneous2025,subotnikAugmentedEhrenfestDynamics2010}. TAB differs from earlier proposals in two ways. First, the decoherence correction is applied pairwise between states, which matters when many states are involved. Second, when combined with the dense-manifold-of-states (DMS) approximation, TAB does not require explicit adiabatic states, enabling efficient on-the-fly dynamics. Analysis of the exact factorization also emphasizes the pairwise nature of decoherence\cite{arribasElectronicCoherencesMolecules2024a}. 

In the TAB method, stochastic collapses result in discontinuous changes in the potential energy, which requires adjusting the momentum of the classical trajectory to conserve total energy. In earlier work in which TAB was applied to low-dimensional models\cite{liangSimulatingPassageCascade} the momentum was scaled isotropically along the momentum direction for convenience.  However, it is well known that this choice is not size-consistent, and can lead to significant errors in higher-dimensional systems. 
The analogous problem in surface hopping has been studied extensively. \cite{tullyTrajectorySurfaceHopping1971,jasperImprovedTreatmentMomentum2003,subotnikNewApproachDecoherence2011,huangFirstPrinciplesDerivation2023, wuLinearAngularMomentum2024,barbattiVelocityAdjustmentSurface2021,sangiogo-gilVelocityRescalingSurface2025,toldoRecommendationsVelocityAdjustment2024, plasserStrongInfluenceDecoherence2019} 
In Tully's original paper on surface hopping \cite{tullyTrajectorySurfaceHopping1971}, he recommended that the momentum be rescaled along the nonadiabatic coupling vector, $\mathbf{d}_{ij}$, but did not provide a rigorous justification for this choice. Since then, many arguments for this choice of direction have been put forth. Tully, and then Coker and Xiao localized the Pechukas force over a hopping timestep to argue that for surface hopping, momentum should be rescaled along the nonadiabatic coupling vector.\cite{tullyMixedQuantumClassical1998, tullyNonadiabaticMolecularDynamics1991, cokerMethodsMolecularDynamics1995,pechukasTimeDependentSemiclassicalScattering1969, pechukasTimeDependentSemiclassicalScattering1969a}

In this paper, we derive an effective nonadiabatic coupling direction for superpositions of states by extending the Pechukas-force argument to superpositions of states.  We also demonstrate how this quantity is easy to calculate when running Ehrenfest dynamics, provided that the final collapsed states are a subset of the initially occupied states. Finally, we propose to rescale momentum in TAB in the a generalized branching plane spanned by $\mathbf{d}_\text{eff}$ and the gradient difference vector $\mathbf{g}_\text{eff}$ between the final and residual states associated with the collapse. We propose rescaling in this branching plane because of a quirk of the TAB method---that collapse events happen significantly \textit{after} population transfer.  It is the direction of the coupling vector at the time of population transfer that is most relevant, and this vector need not align with the coupling vector at the time of decoherence. We test this new rescaling procedure on a linear vibronic coupling (LVC) model of fulvene \cite{gomezBenchmarkingNonadiabaticQuantum2024} and on a 3-state 3-dimensional model simulating passage through two conical intersections. Fulvene has been proposed as a molecular Tully model by Ibele and Curchod\cite{ibeleMolecularPerspectiveTully2020}. In these tests, this prescription is shown to provide better populations and phase space distributions than several reasonable alternatives.

We note that contemporary work by Subotnik and others has emphasized the need to account for angular momentum in a rescale procedure as well\cite{wuLinearAngularMomentum2024,littlejohnRepresentationConservationAngular2023}, but keeping in mind our objective of developing practical and efficient methods for large systems, we limit this paper to only considering the conservation of energy. 

\section{Methods}
\subsection{TAB dynamics}
In this section, we briefly describe the main features of the TAB algorithm. For a detailed description, refer to refs \citenum{eschStatepairwiseDecoherenceTimes2020,eschDecoherencecorrectedEhrenfestMolecular2020,eschAccurateNonempiricalMethod2021}.
As with most mixed quantum-classical trajectory methods, the nuclei are treated as point-like classical particles, while the electrons are treated quantum mechanically. If the electronic system for a classical nuclear trajectory at $\mathbf{R}(t)$  is in some state (not necessarily an eigenstate of the Hamiltonian) $|\psi(\textbf{R},t) \rangle$, the Ehrenfest force acting on the nuclei for this trajectory is
\begin{align}
    \mathbf{F}_\text{Ehr}(\mathbf{R},t)=-\langle\psi(\mathbf{R},t)|\mathbf{\nabla_R} \hat{H}(\mathbf{R}) | \psi(\mathbf{R},t)\rangle_r.
\end{align}
 The subscript $r$ indicates that the expectation value is taken over electronic coordinates only. For the rest of the paper, we drop this subscript and all $\langle \dots \rangle$ denote integrals over electronic coordinates. The nuclei are then propagated classically using this force. 
For each trajectory, we compute the coherent electronic density matrix $\mathbf{\rho}^c(t)=|\psi_\text{initial}(\textbf{R}(t),t)\rangle\langle\psi_\text{initial}(\mathbf{R}(t),t)|$ at each time step (we drop the $\mathbf{R}$ dependence of $\rho$ for brevity). We then scale the off-diagonal elements of $\rho^c (t)$ to form the target (decoherence corrected) density matrix $\rho^d(t)$ according to
\begin{align}
    \label{eq:decoherence correction}
    \rho_{ij}^d(t)&=\rho_{ij}^c(t) e^{\frac{-t}{\tau_{ij}}} \quad\forall i\neq j\\
    \rho_{ij}^d(t)&=\rho_{ij}^c(t) \quad \forall i= j.
\end{align}
Here $\tau_{ij}$ is a state-pairwise decoherence time  based on the work of Bittner and Rossky \cite{bittnerQuantumDecoherenceMixed1995} that depends on the difference of forces between states,

\begin{align}
    \frac{1}{\tau_{ij}}=\sqrt{\sum_N \frac{(F_{ii,N} - F_{jj,N})^2}{8 \hbar^2 \alpha_N}},
\end{align}
where $N$ sums over the nuclear degrees of freedom, and $\alpha_N$ is a constant that needs to be set by the user. In this paper, we set $\alpha_N$ as the width parameter of the initial wavepacket in position space. The forces are diagonal terms of the force in the adiabatic basis 
\begin{align}
    F_{ii,N}(\mathbf{R})=-\langle\phi_i (\mathbf{R})|\frac{\partial}{\partial R_N}H|\phi_i(\mathbf{R})\rangle,
\end{align}
where $|\phi_i(\mathbf{R})\rangle$ is the $i^{th}$ adiabatic state. 

In equation \ref{eq:decoherence correction}, we assume that the loss of coherence follows an exponential decay. Previous work has discussed approximations to a Gaussian decoherence correction, but we limit ourselves to the exponential case for ease of analysis. \cite{liangSimulatingPassageCascade,eschAccurateNonempiricalMethod2021}
TAB now expands $\rho^d(t)$ as a sum of pure state coherent block density matrices
\begin{align}
    \rho^d(t)=\sum_i w_i \rho_i (t).
\end{align}
Refer to previous work for how these blocks are constructed\cite{eschAccurateNonempiricalMethod2021}. To represent decoherence, the weights, $\{w_i\}$, are taken to be the probability of each block, and the wave function collapses to a randomly chosen block, say $\rho_k(t)$. Since $\rho_k(t)$ is a pure density matrix we may write it as an outer product of a final state $|\psi_\text{final}(t)\rangle$
\begin{align}
    \rho_k(t) =|\psi_\text{final}(t)\rangle\langle\psi_\text{final}(t)|.
\end{align}
Again, we emphasize that $|\psi_\text{final}(t)\rangle$ is not necessarily an eigenstate of the Hamiltonian.
We can always write $|\psi_\text{initial}\rangle$ as a linear combination of $|\psi_\text{final}\rangle$ and a `residual' state,  $|\psi_\text{residual}\rangle$. Since there is no phase change from $|\psi_\text{initial}\rangle$ to $|\psi_\text{final}\rangle$ we get,
    \begin{align}
        |\psi_\text{initial}(t)\rangle = \sqrt{P}|\psi_\text{final}(t)\rangle + \sqrt{1-P}|\psi_\text{residual}(t)\rangle,
        \label{eq: psi initial}
    \end{align}
    where $\langle\psi_\text{final}(t)|\psi_\text{residual}(t)\rangle= 0$, and $P$ is real and non-negative.
    
\subsection{The Pechukas force over a collapse step}
\label{section: Pechukas derivation}
Since the collapse event in TAB introduces a discontinuous change in the potential energy, and we require our trajectories to conserve energy, the momentum must be adjusted.  For the analogous problem in surface hopping, analysis of the Pechukas force over a hopping event suggested that the momentum be adjusted along $\mathbf{d}_{ij}$.\cite{cokerMethodsMolecularDynamics1995, tullyMixedQuantumclassicalDynamics1998, tullyNonadiabaticMolecularDynamics1991}
The Pechukas force arises from a stationary phase approximation to the path integral for the reduced nuclear propagator over coupled electronic surfaces.\cite{pechukasTimeDependentSemiclassicalScattering1969, pechukasTimeDependentSemiclassicalScattering1969a,mcwhirterAnalysisPechukasDescription1999} It provides a formally exact effective force on the classical nuclei during an electronic transition, from which the momentum adjustment prescription can be derived.
For a system starting in state $\alpha$ at $t'$ that resolves to state $\beta$ at $t''$, 
the force along a stationary phase path $\mathbf{\bar{R}}$ is
    \begin{align}
        \label{eq:pechukas}
        \mathbf{F}_p(t)=M \mathbf{\ddot{\bar{R}}}(t)=-\text{Re} \bigg\{\frac{\langle\psi_\beta (t,t'')|\nabla_{\mathbf{\bar{R}}(t)}H|\psi_\alpha (t,t')\rangle}{\langle\psi_\beta (t,t'')|\psi_\alpha (t,t')\rangle} \bigg\},
    \end{align}
    where $t'\leq t\leq t''$. The difficulty with developing an algorithm that employs such a force is that it requires advance knowledge of the future state of the system at $t''$, which is then propagated backwards in time to $t$. This makes the force non local and provides a self-consistency criterion.\cite{websterNonadiabaticProcessesCondensed1991}

Suppose that we have some external mechanism to resolve the final state of the system, say surface hopping or TAB. What is the Pechukas force over such a time step?

 There are two main routes that one can take to answer this question.  The first is the argument put forward by Tully\cite{tullyNonadiabaticMolecularDynamics1991}, that calculates the force in equation~\ref{eq:pechukas} by taking the limit $t'' \to t'$. The second is to make a perturbative approximation along the lines of Coker and Xiao.\cite{cokerMethodsMolecularDynamics1995} We choose to take the latter approach and extend the work of Coker and Xiao to arbitrary superpositions of states.
 \par
We analyze the TAB collapse event using the Pechukas force as follows. The collapse is instantaneous at time $t$, $|\psi_\text{initial} (t) \rangle$ is replaced by $|\psi_\text{final}(t) \rangle$, with the decomposition (eq. \ref{eq: psi initial}) holding in a single adiabatic basis $\{|\phi_i (t) \rangle \}$.  To evaluate the Pechukas force at $t$, we consider a fictitious interval $[t,t+\Delta t]$ over which $|\psi_\text{initial} (t) \rangle$ resolves to $|\psi_\text{final}(t+\Delta t)\rangle$.
 At time $t$, the electronic state is
\begin{align}
    |\psi_\text{initial}(t)\rangle = \sum_{i \in \mathcal{I}} a_i(t)\,|\phi_i(t)\rangle.
    \label{eq:initial coeff}
\end{align}
 During the interval $[t, t+\Delta t]$, the state collapses by projection into a final subspace $\mathcal{F} \subset \mathcal{I}$, leaving a residual subspace $\mathcal{R} = \mathcal{I}\setminus\mathcal{F}$. At time $t+\Delta t$, the final and residual states are
\begin{align}
    |\psi_\text{final}(t+\Delta t)\rangle &= \sum_{i \in \mathcal{F}} f_i\,|\phi_i(t+\Delta t)\rangle \\
    |\psi_\text{residual}(t+\Delta t)\rangle &= \sum_{i \in \mathcal{R}} r_i\,|\phi_i(t+\Delta t)\rangle
\end{align}
Practically, in the TAB method, a collapse is instantaneous. This means that we can use eq \ref{eq: psi initial} to decompose $a_i(t)$ as $a_i(t) = \sqrt{P}\,f_i$ for $i\in\mathcal{F}$ and $a_i(t) = \sqrt{1-P}\,r_i$ for $i\in\mathcal{R}$.

The Pechukas force at time $t$ is
\begin{align}
    \mathbf{F}_P(t) = -\text{Re}\bigg\{\frac{\langle\tilde{\psi}_\text{final}(t)|\nabla_{{\mathbf{R}}}H|\psi_\text{initial}(t)\rangle}{\langle\tilde{\psi}_\text{final}(t)|\psi_\text{initial}(t)\rangle}\bigg\},
    \label{eq:pechukas_local}
\end{align}
where $|\tilde{\psi}_\text{final}(t)\rangle = \sum_i c_i(t)\,|\phi_i(t)\rangle$ is obtained by back-propagating $|\psi_\text{final}(t+\Delta t)\rangle$ from $t+\Delta t$ to $t$. 

The boundary conditions for the back-propagation are
\begin{align}
    c_i(t+\Delta t) = \begin{cases}
    f_i & i \in \mathcal{F} \\
    0 & i \in \mathcal{R}
    \end{cases}.
\end{align}
To first order in $\Delta t$, the backward evolution gives
\begin{align}
    c_i(t) = &c_i(t+\Delta t) + \Delta t\left(\frac{iE_i}{\hbar}\,c_i(t+\Delta t) + \sum_k (\mathbf{d}_{ik}\cdot\mathbf{v})\,c_k(t+\Delta t)\right) \nonumber\\
    &+ \mathcal{O}(\Delta t^2),
    \label{eq:final coeff backprop}
\end{align}
where $\mathbf{d}_{ik} = \langle\phi_i|\nabla_\mathbf{R}\phi_k\rangle$ is the nonadiabatic coupling vector between states $i$ and $k$, and $\mathbf{v} = \dot{\mathbf{R}}$ is the nuclear velocity. Since $c_k(t+\Delta t) = 0$ for $k \in \mathcal{R}$, only states in $\mathcal{F}$ contribute to the coupling sum:
\begin{align}
    c_i(t) = \begin{cases}
    f_i\left(1 + \dfrac{i\Delta t\,E_i}{\hbar}\right) + \Delta t\displaystyle\sum_{k\in\mathcal{F}}(\mathbf{d}_{ik}\cdot\mathbf{v})\,f_k & i \in \mathcal{F} \\[10pt]
    \Delta t\displaystyle\sum_{k\in\mathcal{F}}(\mathbf{d}_{ik}\cdot\mathbf{v})\,f_k & i \in \mathcal{R}
    \end{cases}.
    \label{eq:back_prop}
\end{align}
The first-order correction to $c_i$ for $i\in\mathcal{F}$ involves only coupling within $\mathcal{F}$, since $c_k(t+\Delta t) = 0$ for $k\in\mathcal{R}$---the back-propagated state does not couple to $\mathcal{R}$ until second order. We neglect leakage into states outside $\mathcal{I}$, consistent with the Ehrenfest approximation.
We now expand the denominator, $\mathcal{D}$, in equation \ref{eq:pechukas_local} with the adiabatic coefficients from equations \ref{eq:initial coeff} and \ref{eq:final coeff backprop}.
\begin{align}
    \mathcal{D} &=\langle\tilde{\psi}_\text{final}(t)|\psi_\text{initial}(t)\rangle = \sum_{i \in \mathcal{I}}c^*_i(t)a_i (t) \\
     \mathcal{D} &= \sum_{i\in\mathcal{F}} c_i^*(t)\,\sqrt{P}\,f_i + \sum_{i\in\mathcal{R}} c_i^*(t)\,\sqrt{1-P}\,r_i \\
    \mathcal{D} &= \sqrt{P}\left[\sum_{i\in\mathcal{F}}|f_i|^2\left(1 - \frac{i\Delta t\,E_i}{\hbar}\right) + \Delta t\sum_{i,k\in\mathcal{F}}(\mathbf{d}_{ik}\cdot\mathbf{v})^*f_k^*f_i\right] \nonumber\\
    &\quad + \sqrt{1-P}\,\Delta t\sum_{i\in\mathcal{R}}\sum_{k\in\mathcal{F}}(\mathbf{d}_{ik}\cdot\mathbf{v})^*f_k^*\,r_i.
    \label{eq:denominator}
\end{align}
Using the fact that $\sum_{i\in\mathcal{F}}|f_i|^2 = 1$, and collecting all the $\Delta t$ terms as $\mathcal{D'}$, we rewrite equation \ref{eq:denominator} as 
\begin{align}
    &\mathcal{D}=\sqrt{P}+\Delta t\mathcal{D'}.
\end{align}
We further denote the real and imaginary parts of $\mathcal{D'}$ as $\mathcal{D'}_\text{Re}=\text{Re}\{\mathcal{D'}\}$ and $\mathcal{D'}_\text{Im}=\text{Im}\{\mathcal{D'}\}$ respectively. 
Now, the numerator of equation \ref{eq:pechukas_local},
    \begin{align}
    \mathcal{N} &= \sum_{i,j \in \mathcal{I}} c_i^*(t)\,a_j(t)\,\langle\phi_i|\nabla_{ {\mathbf{R}}}H|\phi_j\rangle.
\end{align}
We split this sum according to the subspace membership of $i$ and
$j$:
\begin{align}
    \mathcal{N} &= \underbrace{\sum_{i,j \in \mathcal{F}} c_i^* a_j \langle\phi_i|\nabla_{ {\mathbf{R}}}H|\phi_j\rangle}_{\mathcal{N}_\mathcal{FF}} \nonumber
    \quad + \underbrace{\sum_{\substack{i \in \mathcal{F} \\ j \in \mathcal{R}}} c_i^* a_j \langle\phi_i|\nabla_{ {\mathbf{R}}}H|\phi_j\rangle}_{\mathcal{N}_\mathcal{FR}} \nonumber\\
    &\quad + \underbrace{\sum_{\substack{i \in \mathcal{R} \\ j \in \mathcal{F}}} c_i^* a_j \langle\phi_i|\nabla_{ {\mathbf{R}}}H|\phi_j\rangle}_{\mathcal{N}_\mathcal{RF}}
    \quad + \underbrace{\sum_{i,j \in \mathcal{R}} c_i^* a_j \langle\phi_i|\nabla_{ {\mathbf{R}}}H|\phi_j\rangle}_{\mathcal{N}_\mathcal{RR}}.
\end{align}
Using the Hellmann--Feynman relations
\begin{align}
    \langle\phi_i|\nabla_{{\mathbf{R}}}H|\phi_j\rangle = \begin{cases} \nabla_{{\mathbf{R}}} E_i & i=j \\ (E_j - E_i)\,\mathbf{d}_{ij} & i \neq j \end{cases},
\end{align}
and noting that $i\neq j$ for all terms in $\mathcal{N}_{\mathcal{FR}}$ and $\mathcal{N}_{\mathcal{RF}}$, we substitute the back-propagated coefficients. 
\begin{align}
    \mathcal{N}_{\mathcal{FF}} &= \sqrt{P}\sum_{i,j\in\mathcal{F}}\left(f_i \bigg(1+ \tfrac{i\Delta t\,E_i}{\hbar} \bigg)+ \Delta t\sum_{k\in\mathcal{F}}(\mathbf{d}_{ik}\cdot\mathbf{v})\,f_k\right)^*f_j\,\langle\phi_i|\nabla_{{\mathbf{R}}}H|\phi_j\rangle,
\end{align}
\begin{align}
    \mathcal{N}_{\mathcal{FR}}&= \sqrt{1-P}\sum_{\substack{i\in\mathcal{F}\\j\in\mathcal{R}}}\left(f_i \bigg(1+ \tfrac{i\Delta t\,E_i}{\hbar} \bigg)+\Delta t\sum_{k\in\mathcal{F}}(\mathbf{d}_{ik}\cdot\mathbf{v})\,f_k\right)^*r_j\,(E_j - E_i)\,\mathbf{d}_{ij},
\end{align}
\begin{align}
    \mathcal{N}_{\mathcal{RF}} &= \sqrt{P}\,\Delta t\sum_{\substack{i\in\mathcal{R}\\j\in\mathcal{F}}}\left(\sum_{k\in\mathcal{F}}(\mathbf{d}_{ik}\cdot\mathbf{v})\,f_k\right)^*f_j\,(E_j - E_i)\,\mathbf{d}_{ij},
\end{align}
\begin{align}
    \mathcal{N}_{\mathcal{RR}} &= \sqrt{1-P}\,\Delta t\sum_{i,j\in\mathcal{R}}\left(\sum_{k\in\mathcal{F}}(\mathbf{d}_{ik}\cdot\mathbf{v})\,f_k\right)^*r_j\,\langle\phi_i|\nabla_{{\mathbf{R}}}H|\phi_j\rangle.
\end{align}
$\mathcal{N}_{\mathcal{FF}}$ and $\mathcal{N}_{\mathcal{FR}}$ contain $\mathcal{O}(1)$ terms, while $\mathcal{N}_{\mathcal{RF}}$ and $\mathcal{N}_{\mathcal{RR}}$ are entirely $\mathcal{O}(\Delta t)$, arising from the back-propagation leakage of $\langle\tilde{\psi}_\text{final}|$ into the residual subspace. 
Identifying the zeroth-order contributions, we define the Ehrenfest force on the final subspace
\begin{align}
    \mathbf{F}_f &= \sum_{i,j\in\mathcal{F}} f^*_i\,f_j\,\langle\phi_i|\nabla_{{\mathbf{R}}}H|\phi_j\rangle \\
    \mathbf{F}_f&= \sum_{i\in\mathcal{F}}|f_i|^2\nabla_{{\mathbf{R}}} E_i + \sum_{\substack{i,j\in\mathcal{F}\\i\neq j}}f^*_i\,f_j\,(E_j-E_i)\,\mathbf{d}_{ij},
\end{align}
and the real component of the inter-subspace coupling vector
\begin{align}
    \mathbf{d}_\text{eff} = \text{Re} \left\{\sum_{\substack{i\in\mathcal{F}\\j\in\mathcal{R}}} f^*_i\,r_j\,(E_j - E_i)\,\mathbf{d}_{ij} \right\}.
\end{align}
The full Pechukas force, keeping all terms to first order in $\Delta t$, is then
\begin{align}
    \mathbf{F}_P(t) &\approx -\mathbf{F}_f - \sqrt{\frac{1-P}{P}}\mathbf{d}_\text{eff}-\frac{\Delta t}{P} \bigg[ P\,(\delta\mathbf{F}_f + \delta\mathbf{d}'_\text{eff})+\\ \nonumber 
    &\sqrt{P(1-P)}(\delta\mathbf{d}_\text{eff} + \delta\mathbf{F}_r) -\mathcal{D'}_{Re}(\sqrt{P}\mathbf{F}_f+\sqrt{1-P}\mathbf{d}_\text{eff})\\ \nonumber&+\sqrt{P}\text{Im}\{\mathbf{F_\phi}\}+\sqrt{1-P}\mathbf{b}_\text{eff} \mathcal{D'}_{Im}\bigg],
    \label{eq:full_pechukas}
\end{align}
where the $\mathcal{O}(\Delta t)$ corrections are
\begin{align}
    \mathbf{b}_\text{eff}&=\text{Im}\bigg\{\sum_{\substack{i\in\mathcal{F}\\j\in\mathcal{R}}} f^*_i\,r_j\,(E_j - E_i)\,\mathbf{d}_{ij} \bigg\},\\
    \delta\mathbf{F}_f &=\text{Re} \bigg\{\sum_{i,j,k\in\mathcal{F}}(\mathbf{d}_{ik}\cdot\mathbf{v})^*\,f_k^*\,f_j\,\langle\phi_i|\nabla_{{\mathbf{R}}}H|\phi_j\rangle\bigg\}\\
    \delta\mathbf{d}_\text{eff} &=  \text{Re} \bigg\{\sum_{\substack{i,k\in\mathcal{F}\\j\in\mathcal{R}}}(\mathbf{d}_{ik}\cdot\mathbf{v})^*\,f_k^*\,r_j\,(E_j-E_i)\,\mathbf{d}_{ij}\bigg\}\\
    \delta\mathbf{d}'_\text{eff} &= \text{Re} \bigg\{\sum_{\substack{i\in\mathcal{R},\,k\in\mathcal{F}\\j\in\mathcal{F}}}(\mathbf{d}_{ik}\cdot\mathbf{v})^*\,f_k^*\,f_j\,(E_j-E_i)\,\mathbf{d}_{ij}\bigg\}\\
    \delta\mathbf{F}_r &= \text{Re} \bigg\{\sum_{\substack{i\in\mathcal{R},\,k\in\mathcal{F}\\j\in\mathcal{R}}}(\mathbf{d}_{ik}\cdot\mathbf{v})^*\,f_k^*\,r_j\,\langle\phi_i|\nabla_{{\mathbf{R}}}H|\phi_j\rangle\bigg\},\\
    \mathbf{F_\phi}&= \sum_{i,j \in \mathcal{F}}\tfrac{\sqrt{P}f_i ^* f_jE_i}{\hbar}\langle\phi_i|\nabla_{{\mathbf{R}}}H|\phi_j\rangle+ \sum_{{\substack{i\in\mathcal{F}\\j\in\mathcal{R}}}}\tfrac{\sqrt{1-P}f_i ^* r_jE_i}{\hbar}(E_j - E_i)\,\mathbf{d}_{ij} 
\end{align}
The final expression (equation \ref{eq:full_pechukas}) has an interesting structure. The first term, $-{\mathbf{F}_f}$, is the existing Ehrenfest force at the end of the TAB collapse. The other $\mathcal{O}(1)$ term is the effective nonadiabatic coupling vector, which is weighted by the energy gap between states. At first glance, this should be the direction along which the momentum should be adjusted. Among the first order terms, $\delta \mathbf{F}_f$ and $\delta \mathbf{F}_r$ are first order corrections to the final and left-behind residual Ehrenfest forces respectively. The first order terms $\delta \mathbf{d}_\text{eff}$ and $\delta \mathbf{d}'_\text{eff}$ are corrections to $\mathbf{d}_\text{eff}$. 
The other first order term, $\mathbf{F}_\phi$, is a phase correction to Ehrenfest.
We also note that in the case that a collapse may not happen, i.e, $P=1$, the leading term of the Pechukas force is simply the Ehrenfest force.

\subsection{Rescaling algorithms}
In this work we test three algorithms for rescaling the momentum: Isotropic momentum rescaling, rescaling along the effecting nonadiabatic coupling vector, $\mathbf{d}_\text{eff}$, and rescaling within the branching plane.  Here we define these three approaches.
    
To conserve energy over a collapse time step, we write,
\begin{align}
    \langle\psi_\text{initial}|\hat{H}|\psi_\text{initial}\rangle + \sum_\eta \frac{\mathbf{p_\eta}^2}{2M_\eta} = \langle\psi_\text{final}|\hat{H}|\psi_\text{final}\rangle + \sum_\eta \frac{|\mathbf{p_\eta} + \gamma \mathbf{u_\eta} |^2}{2M_\eta},
    \label{eq:energy cons}
\end{align}
where $\eta$ indexes the nuclei with $\mathbf{p}_\eta$ the three-dimensional momentum of nucleus $\eta$.  The three rescaling algorithms tested in this work are differentiated by the definition of $\mathbf{u}$

This formulation is based on two approximations.  The first approximation is that we neglect angular momentum.\cite{wuLinearAngularMomentum2024,littlejohnRepresentationConservationAngular2023}  
The second approximation is that we enforce energy conservation on a per-trajectory basis.\cite{martensSurfaceHoppingConsensus2016,ibeleCoupledTrajectoryStrategyDecoherence2026} We think these approximations are reasonable given our goal of developing a fast and efficient algorithm for large systems. 

Equation \ref{eq:energy cons} gives a quadratic equation for $\gamma$ whose solutions are
\begin{align}
\gamma = \frac{-b \pm \sqrt{b^2 - 4a\Delta E}}{2a},
\label{eq:gamma solution}
\end{align}
where $a=\sum_\eta \frac{\mathbf{u_\eta}\cdot\mathbf{u_\eta}}{2M_\eta}$,$b= \sum_\eta \frac{\mathbf{p_\eta}\cdot\mathbf{u_\eta}}{M_\eta}$ and $\Delta E = \langle\psi_\text{final}|\hat{H}|\psi_\text{final}\rangle-\langle\psi_\text{initial}|\hat{H}|\psi_\text{initial}\rangle$. We choose the solution to $\gamma$ in equation \ref{eq:gamma solution} that has a smaller magnitude, as is common in surface hopping.\cite{barbattiVelocityAdjustmentSurface2021,fabianoImplementationSurfaceHopping2008}  A collapse is frustrated if equation \ref{eq:gamma solution} has no real solution. In this case, the $\mathbf{u}$ component of the momentum is reversed. 

\subsubsection{Isotropic rescaling along $p$}
In past work using the TAB algorithm, the momentum has been scaled, that is, $\mathbf{u}=\mathbf{p}/|\mathbf{p}|$. As has been pointed out in the literature, such a momentum rescaling procedure is not size consistent. Hereafter, we refer to this algorithm as ``rescaling along $p$,'' or simply ``the $p$ method.''

\subsubsection{Rescaling along $\mathbf{d}_\text{eff}$}
Building off the above formalism, we argue that scaling in the $\mathbf{d}_\text{eff}$ direction is physically well-motivated.
The zeroth-order terms of eq \ref{eq:full_pechukas} are $-\mathbf F_f-\sqrt{(1-P)/P}\,\mathbf d_\text{eff}$. Of the two terms, $-\mathbf F_f$ is the Ehrenfest force on $|\psi_\text{final}\rangle$, which the propagation already applies following the collapse; the contribution unique to the collapse event is the term $-\sqrt{(1-P)/P}\,\mathbf d_\text{eff}$.  Thus, we test rescaling along this direction, $\mathbf{u}=\hat{\mathbf{d}}_{\text{eff}}=\mathbf{d}_{\text{eff}}/|\mathbf{d}_{\text{eff}}|$. We call $\mathbf{d}_{\text{eff}}$ the effective nonadiabatic coupling vector by analogy with $\mathbf{d}_{ij}$ in the two-state case, noting that $\mathbf{d}_{\text{eff}}$ carries an additional energy-gap weighting and therefore has units of force rather than inverse length. The normalized rescaling direction is unaffected by this distinction.
One challenge in practical implementations of surface hopping is that developing analytic calculations of the NAC vector requires coding effort that it would be desirable to avoid.\cite{barbattiVelocityAdjustmentSurface2021,fabianoImplementationSurfaceHopping2008,hammes-schifferProtonTransferSolution1994,meekEvaluationTimeDerivativeCoupling2014}.  

Here we are able to efficiently calculate $\mathbf{d}_{\text{eff}}$ from a set of mean-field force calculations without explicitly calculating the nonadiabatic couplings
\begin{align}
\label{eq: deff calc}
\mathbf{d}_\text{eff}=&\frac{1}{2\sqrt{P(1-P)}}\bigg[\langle\psi_\text{initial}|\nabla H|\psi_\text{initial}\rangle-P\langle\psi_\text{final}|\nabla H|\psi_\text{final}\rangle\\ \nonumber
&-(1-P)\langle\psi_\text{residual}|\nabla H|\psi_\text{residual}\rangle\bigg].
\end{align}

\subsubsection{Rescaling in the branching plane}
The Pechukas analysis in section \ref{section: Pechukas derivation} assumed that the collapse event is local in time. While the TAB algorithm enforces this locality at the level of the algorithm (the collapse happens instantaneously at a single timestep), the physical decoherence (equation \ref{eq:decoherence correction}) is continuous: coherence between states is lost gradually as their forces differ. The collapse instant in TAB is determined stochastically once enough population has transferred between states.  
Ideally, one would evaluate the non-local Pechukas force over the full coherence-decay window, as Webster, Rossky, and Friesner advocate for surface hopping.\cite{websterNonadiabaticProcessesCondensed1991} The first-order terms in equation \ref{eq:full_pechukas} represent the leading contribution of this non-local force. Evaluating them on the fly, however, would require explicit nonadiabatic couplings and, because collapses in TAB are stochastic, even the exact first-order force would be evaluated at an essentially arbitrary point within the coherence-decay window. We therefore retain only the zeroth-order terms.  The dominant neglected effect is the rotation of $\mathbf d_\text{eff}$ between the time of population transfer and the collapse instant; this rotation is confined to the branching plane (exactly so in the two-state case). The population transfer itself occurred when the classical momentum was most aligned with the coupling vector.  

In the case of a system with only two-states, one can envision a conical intersection. When a trajectory passes near a conical intersection, the momentum likely varies much more slowly than the coupling direction (which varies discontinuously at the intersection point itself). Thus, it retains the direction of the coupling at the time of transfer (see Fig. \ref{fig:branching plane example}). With this picture in mind, we therefore propose to rescale the component of the momentum that lies in the branching plane.
We define the gradient difference vector $\mathbf{g}_\text{eff}$,
\begin{align}
    \mathbf{g}_\text{eff}=\langle\psi_\text{final}|\nabla\hat{H}|\psi_\text{final}\rangle-\langle\psi_\text{initial}|\nabla\hat{H}|\psi_\text{initial}\rangle\bigg.
\end{align}
We adjust the momentum in the plane spanned by $\mathbf{d}_\text{eff}$ and $\mathbf{g}_\text{eff}$
\begin{align}
    \mathbf{u}= (\mathbf{p}\cdot \hat{\mathbf{\theta}}_{g})\hat{\mathbf{\theta}}_{g} +(\mathbf{p}\cdot \hat{\mathbf{\theta}}_{d})\hat{\mathbf{\theta}}_{d}, 
    \label{eq:branching plane direction}
\end{align}
where $\hat{\mathbf{\theta}}_{d}=\mathbf{d}_{\text{eff}}/|\mathbf{d}_{\text{eff}}|$ and 
\begin{align}
  \hat{\mathbf{\theta}}_{g}=\frac{\mathbf{g}_{\text{eff}}-(\mathbf{g}_{\text{eff}} \cdot\hat{\mathbf{\theta}}_{d})\hat{\mathbf{\theta}}_{d}}{|\mathbf{g}_{\text{eff}}-(\mathbf{g}_{\text{eff}} \cdot\hat{\mathbf{\theta}}_{d})\hat{\mathbf{\theta}}_{d}|}  
\end{align}
 We label this rescaling choice as `branching plane' in the results. Note that the residual and final states are not necessarily adiabatic electronic states, thus we use the term branching plane in a more general sense here than is usual.  In this work, the branching plane is defined by the nonadiabatic coupling vector and the difference gradient direction between the final and residual states, regardless of whether they are adiabatic states or not.
\section{Computational Details}

\subsection{Fulvene LVC model}

We test our methods on an LVC model of fulvene.\cite{gomezBenchmarkingNonadiabaticQuantum2024} The S1-S0 ultrafast conversion of fulvene is one of the molecular Tully models proposed by Ibele and Curchod.\cite{ibeleMolecularPerspectiveTully2020} Gómez et al.\cite{gomezBenchmarkingNonadiabaticQuantum2024} parametrized an LVC model containing all 30 vibrational degrees of freedom. We compare our TAB results to the MCTDH results from the same work. This model has recently been used to study the effect of momentum rescaling in surface hopping approaches.\cite{mannouchQuantumQualityClassical2024,schurgerAssessingPerformanceCoupledtrajectory2025} 2000 TAB trajectories were run for each momentum rescaling choice. We chose the initial conditions to match the MCTDH calculations, namely, the initial trajectories were sampled from the Wigner transform in positions and momenta of the S0 vibrational harmonic oscillator ground state. This ensemble wavepacket was then instantaneously vertically excited to the S1 state. At the Franck-Condon point, the adiabatic state (S1) coincides with diabatic state 1. 
\subsection{3-State Model}

\label{section: 3 state model}
To test our method further, we use a 3-state, 3 dimensional model inspired from ref \citenum{liangSimulatingPassageCascade}. It replicates passage through consecutive conical intersections (CI).\cite{domckeRoleConicalIntersections2012} We define the Hamiltonian in the diabatic basis as
\begin{align}
    H(x_1,x_2)=\begin{pmatrix}
0.25x_1 & 0.025x_2 & 0.025x_2\\
0.025x_2 & -0.025x_1 & 0 \\
0.025x_2 & 0 & -0.025x_1 - 0.01
\end{pmatrix}
\end{align}
where all numbers are in atomic units. The model comprises a tuning mode, $x_1$, and an off-diagonal coupling mode, $x_2$ that couples states 1 and 2 to state 0. We also add a spectator mode to the system, $x_3$. The effect of spectator modes has been studied in the context of other nonadiabatic methods.\cite{worthEffectModelEnvironment1996, pereiraQuantumMolecularDynamics2023,salaQuantumDynamicsMultidimensional2018} In our case, the Hamiltonian is independent of $x_3$, which implies that the dynamics of the system should also be independent of this mode. We ran an ensemble of 2000 TAB trajectories for each momentum rescaling choice of direction. The trajectories were initialized on diabatic state 0. Classical positions and momenta were sampled from the Wigner transform of a Gaussian wavepacket centered at $x_1=-1, \; x_2=0,\; x_3=0$ and $p_1=10,\; p_2=10,\; p_3=10$. The widths of the gaussian wavepacket were the same for all coordinates with $\sigma_x= 0.204,\; \sigma_p=2.451.$ The mass of the nucleus is set to 1845 a.u. Numerically exact results were computed by propagating the time-dependent Schrödinger equation on a uniform $1000^3$ Cartesian grid spanning $x_1,x_2,x_3 \in [-4, 8]$ a.u., with the kinetic-energy operator discretized by a 4th-order central finite-difference Laplacian and the diabatic potential plus interstate couplings applied pointwise. Time integration used a second-order symplectic split-operator scheme in which the real and imaginary parts of the wavefunction are propagated on staggered half-steps.\cite{visscherFastExplicitAlgorithm1991}

\section{Results}

\begin{figure}
    \centering
    \includegraphics[width=8cm]{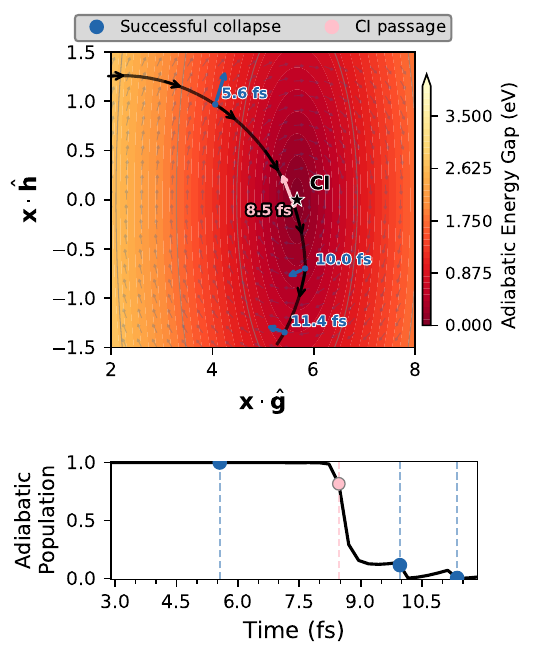}
    \caption{The upper panel shows an excerpt of a single TAB trajectory's passage through the CI of the fulvene LVC model in the branching plane. The lower panel shows the adiabatic population for this trajectory. The successful collapse events are marked in blue. The blue arrows in the upper panel point in the direction of $\mathbf{d}_\text{eff}.$ The vector field in the background is the analytically calculated $\mathbf{d}_{01}$ field. The pink arrow shows the direction of $\mathbf{d}_{01}$ at the point where the trajectory is closest to the CI.}
    \label{fig:branching plane example}
\end{figure}
We compare three momentum rescaling prescriptions: rescaling along the instantaneous $\hat{\mathbf{d}}_{\text{eff}}$ ($\mathbf{u} = \hat{\mathbf{d}}_{\text{eff}}$), isotropic rescaling along $\mathbf{p}$ ($\mathbf{u} = \hat{\mathbf{p}}$, the prescription used in earlier TAB work\cite{eschAccurateNonempiricalMethod2021}), and rescaling in the branching plane (equation \ref{eq:branching plane direction}). 

Figure \ref{fig:branching plane example} demonstrates the issue of rescaling momentum along $\mathbf{d}_\text{eff}$ by showing the dynamics in the branching plane for a single TAB trajectory for the fulvene model. The upper panel shows the trajectory in the branching plane spanned by the diabatic gradient difference vector $\hat{\mathbf{g}}$ and the diabatic coupling vector $\hat{\mathbf{h}}$. The background vector field is the analytically computed NAC, $\hat{\mathbf{d}}_{01}$,
 which varies rapidly in the neighborhood of the CI. The blue arrows in the upper panel show the computed $\mathbf{d}_\text{eff}$ at successful collapse events. Since this model has only two states, $\mathbf{d}_\text{eff}$ and $\hat{\mathbf{d}}_{01}$ point in the same direction.

 The lower panel in figure \ref{fig:branching plane example} shows the corresponding adiabatic population for the trajectory.
The first collapse event (labeled 1) happens before reaching the CI where the trajectory collapses to the upper adiabatic state. The collapses to the lower state (labeled 2-6) happen much after passing through the CI, after population transfers and coherence is lost. However, the direction of $\mathbf{d}_\text{eff}$ at collapse 2 is not the only force direction that contributed to the collapse event. The impulse acting on the trajectory over the coherence-decay window samples $\mathbf{d}_{\text{eff}}$ at multiple geometries, including the near-CI region where $\mathbf{d}_{\text{eff}}$ was aligned with the momentum. The instantaneous $\mathbf{d}_{\text{eff}}$ at the collapse instant is one snapshot of this average.

\begin{figure}
    \centering
    \includegraphics[width=8cm]{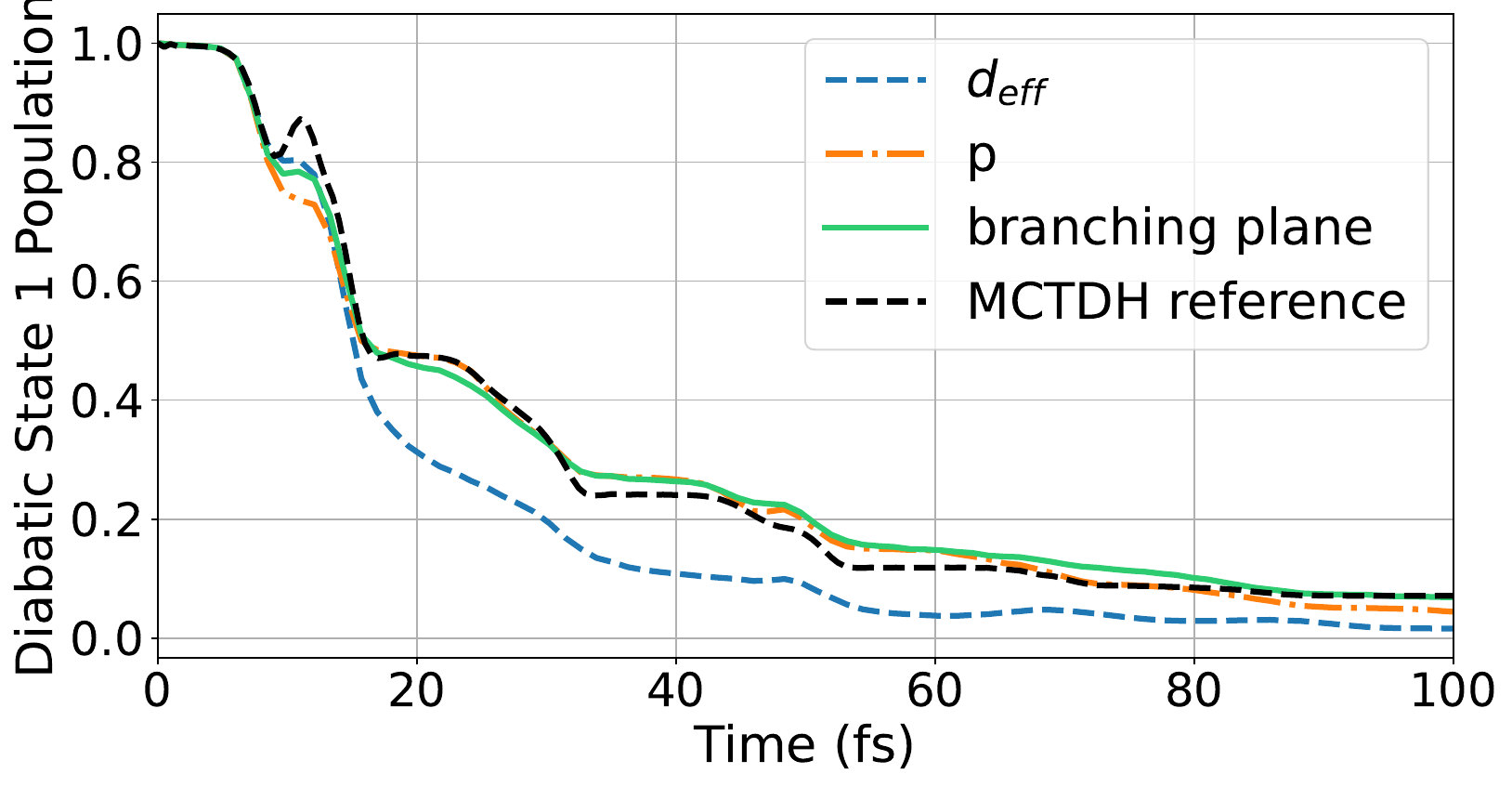}
    \caption{Ensemble diabatic populations of the first state for the fulvene LVC model. We see that rescaling momentum along $\mathbf{d}_\text{eff}$ does worse than the other methods. The MCTDH results were taken from ref \citenum{gomezBenchmarkingNonadiabaticQuantum2024}.}
    \label{fig:fulvene pop}
\end{figure}

Figure \ref{fig:fulvene pop} shows the diabatic populations for state 1 of  the fulvene model. The branching-plane and isotropic prescriptions both reproduce the MCTDH population closely, with the branching-plane result tracking the benchmark marginally more accurately at long times. Rescaling along $\hat{\mathbf{d}}_{\text{eff}}$ produces a population that decays substantially faster than the reference. However, both $\mathbf{d}_\text{eff}$ and branching plane methods do better than $\mathbf{p}$ at capturing the small peak in population before 20 fs.
We note that the fulvene LVC model is a two-state system. In two states, $\mathbf{d}_{\text{eff}} = (E_1 - E_0)\mathbf{d}_{01}$, so the normalized rescaling direction $\hat{\mathbf{d}}_{\text{eff}}$ coincides with the standard nonadiabatic coupling direction $\hat{\mathbf{d}}_{01}$ --- the energy-gap weighting is a scalar prefactor that vanishes under normalization. 

\begin{figure}
    \centering
    \includegraphics[width=8cm]{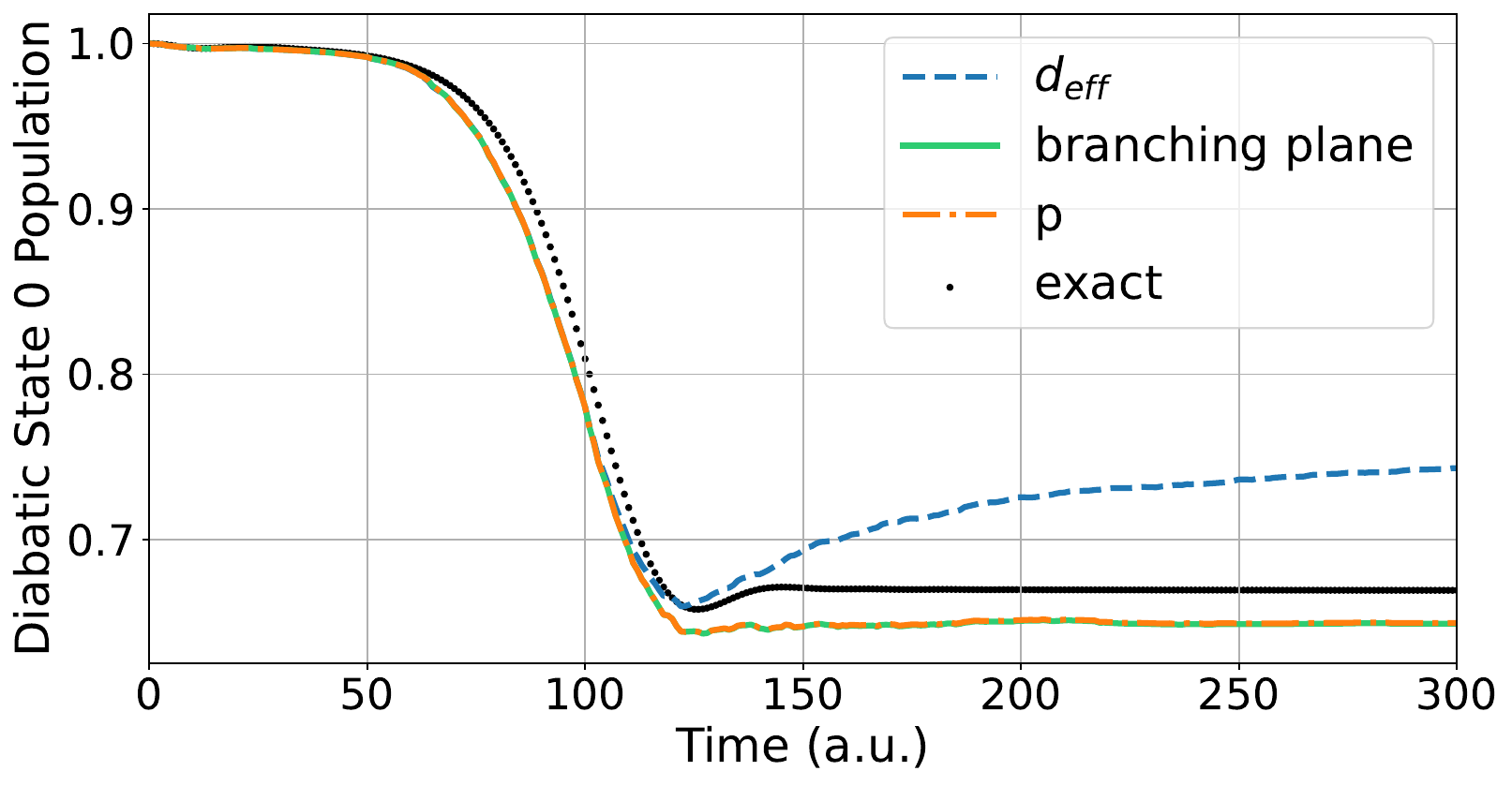}
    \caption{Ensemble diabatic populations for the 3-state model.}
    \label{fig:m3 pop}
\end{figure}

To test our momentum rescaling  prescription on more than 2 states, we study the three rescaling choices on the 3-state model described in section \ref{section: 3 state model}.
Figure \ref{fig:m3 pop} shows diabatic populations for state 0. Both branching plane and $\mathbf{p}$ track the numerically exact population closely. However, $\hat{\mathbf{d}}_\text{eff}$ pushes population back into state 0, beginning at around $t=130$ a.u.

\begin{figure*}[h]
    \centering
    \includegraphics[height=4cm]{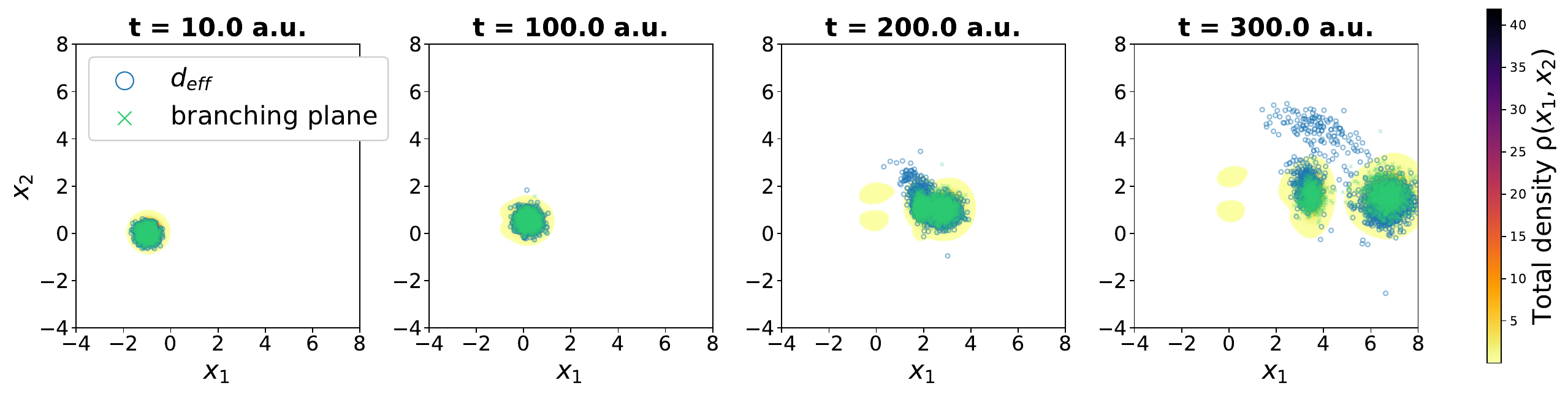}
    \caption{Snapshots of the trajectories in x1 and x2 at different times. The classical trajectories are overlaid on top of the numerically exact marginal density. We see that rescaling along $\mathbf{d}_\text{eff}$ unphysically pushes trajectories along $x_2.$}
    \label{fig:m3 position}
\end{figure*}

We explain the poor performance of $\hat{\mathbf{d}}_\text{eff}$ by examining the position space distributions of the tuning ($x_1$) and coupling ($x_2$) modes. Figure \ref{fig:m3 position} shows the position-space distribution of the classical trajectories at four times, overlaid on the marginal density from the exact grid calculation. At $t=10$ a.u. and $t = 100$ a.u., both rescaling prescriptions track the exact density. By $t=200$ a.u., trajectories rescaled along $\hat{\mathbf{d}}_{\text{eff}}$ have been driven anomalously upward in $x_2$, spreading along this axis far beyond the exact distribution. By $t = 300$ a.u., the $\hat{\mathbf{d}}_{\text{eff}}$ trajectories occupy regions of phase space with negligible exact density, while the branching-plane trajectories remain co-located with the exact density. 

\begin{figure}
    \centering
    \includegraphics[width=8cm]{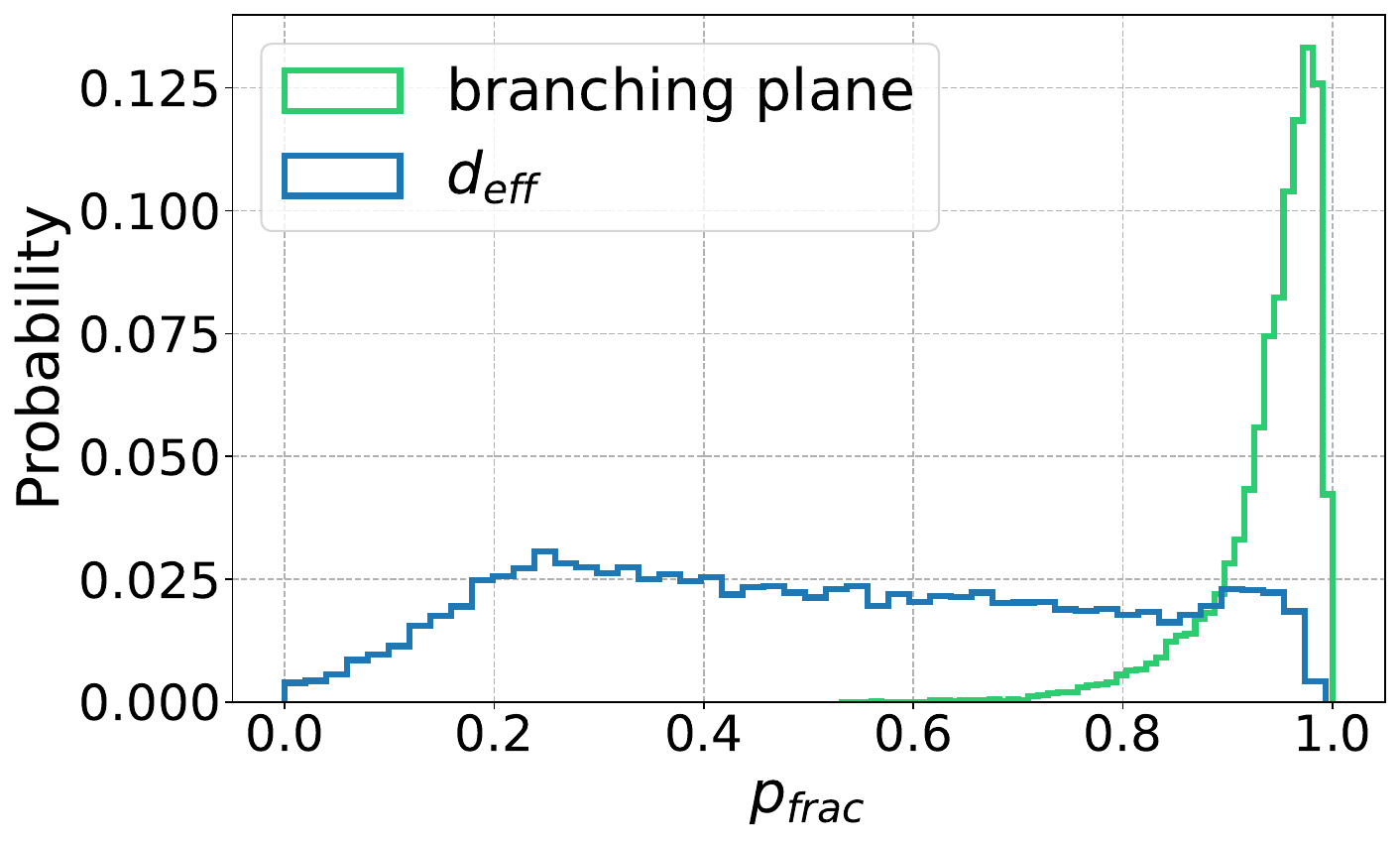}
    \caption{The probability distribution of the momentum along the rescaling vector at a collapse event. $p_\text{frac}=\frac{\mathbf{p \cdot u}}{\mathbf{|p||u|}}$ where $\mathbf{u}$ is the rescaling direction.}
    \label{fig:pfrac m3}
\end{figure}

One observation from the simulations is that the poor overlap between $\mathbf{p}$ and $\hat{\mathbf{d}}_{\text{eff}}$  causes frustrated collapses when rescaling along $\hat{\mathbf{d}}_{\text{eff}}$. Figure \ref{fig:pfrac m3} plots the distribution of the overlap between the momentum and the rescaling vector direction at the time of a collapse. Because the collapse event can happen after passing through a CI,  the $\hat{\mathbf{d}}_{\text{eff}}$ has rotated sufficiently such that it is no longer the physically motivated coupling direction. In contrast, the fraction of the momentum in the branching plane remains large enough to allow for successful collapses.

\begin{figure*}
    \centering
    \includegraphics[width=15cm]{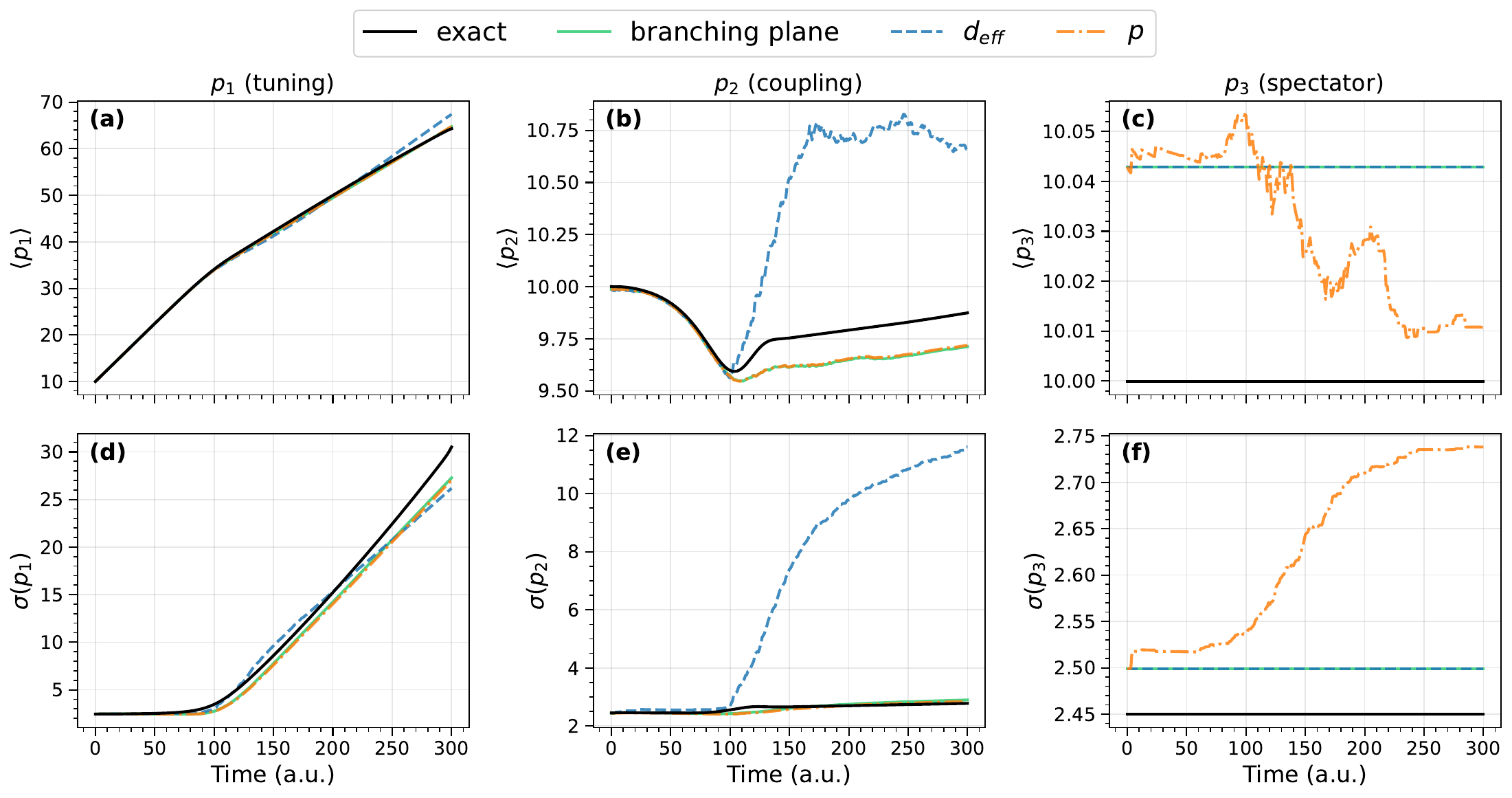}
    \caption{Average momentum (upper panel), and momentum standard deviation(lower panel) for the 3 state model. The fixed error between classical and exact results in $p_3$ is due to sampling. $\mathbf{d}_\text{eff}$ adds extra momentum into the coupling mode, while $\mathbf{p}$ unphysically adds to the spectator mode. Adjusting in the branching plane gives the best results.}
    \label{fig:m3 momentum}
\end{figure*}

One may be tempted to believe that rescaling along $\mathbf{p}$ is good enough, but we emphasize that populations can be a coarse description of the accuracy of a method. Figure \ref{fig:m3 momentum} examines the momentum statistics for the 3-state model. The upper panels (a-c) show the average momentum along the 3 modes. The lower 3 panels (d-f) show the standard deviation of the momentum. Corroborating the position-space distributions of figure \ref{fig:m3 position}, rescaling along $\hat{\mathbf{d}}_{\text{eff}}$ pumps momentum into the coupling mode ($p_2$). Although the Hamiltonian is independent of the spectator mode, $p_3$, rescaling along $\mathbf{p}$ pumps energy into this mode, unphysically (figures \ref{fig:m3 momentum}c and \ref{fig:m3 momentum}f). We note that the fixed error between the numerically exact and the averaged TAB trajectories momenta comes from sampling the Wigner distribution with a finite number of trajectories. 
\section{Conclusion}

We have derived an effective nonadiabatic coupling vector $\mathbf{d}_{\text{eff}}$ for collapses involving superpositions of states by extending the Pechukas-force argument of Coker and Xiao\cite{cokerMethodsMolecularDynamics1995} to arbitrary final subspaces. We find the leading, additional term of the Pechukas force at a collapse instant to be
\begin{equation}
\mathbf{d}_{\text{eff}} = \text{Re}\bigg\{\sum_{\substack{i \in F \\ j \in R}} f^*_i\, r_j\, (E_j - E_i)\, \mathbf{d}_{ij}\bigg\}.
\end{equation}
Interestingly, the energy weights that arise in this Pechukas force do not arise in the expression that governs population transfer,
\begin{equation}
\mathbf{d}^\text{old}_{\text{eff}} = \text{Re}\bigg\{\sum_{\substack{i \in F \\ j \in R}} f^*_i\, r_j\, \mathbf{d}_{ij}\bigg\},
\end{equation}
which was proposed in earlier work on stochastic Ehrenfest and surface hopping methods\cite{bedard-hearnMeanfieldDynamicsStochastic2005,tomazEhrenfestDynamicsSpontaneous2025,subotnikNewApproachDecoherence2011}. 
The distinction is invisible for two-state systems, where the gap appears as a scalar prefactor that vanishes under normalization, but is meaningful whenever multiple state pairs contribute with different gaps. We further show that $\mathbf{d}_{\text{eff}}$ can be evaluated in TAB without explicit nonadiabatic coupling calculations, requiring only one additional force evaluation on the residual state.

For the TAB method specifically, we find that rescaling along the instantaneous $\hat{\mathbf{d}}_{\text{eff}}$ alone produces poor populations and phase-space distributions despite $\mathbf{d}_{\text{eff}}$ being the formally correct impulsive direction at any instantaneous nonadiabatic event. This discrepancy arises because TAB collapse events occur stochastically once coherence has decayed, by which time the trajectory has likely moved past the region of strong nonadiabatic coupling, and $\mathbf{d}_{\text{eff}}$ has rotated within the branching plane. We therefore adjust momentum in the plane spanned by $\mathbf{d}_{\text{eff}}$ and the gradient-difference vector $\mathbf{g}_{\text{eff}}$. This prescription reproduces benchmark populations on both test systems and eliminates the anomalous coupling-mode excitation seen with $\hat{\mathbf{d}}_{\text{eff}}$ rescaling.  In addition, this approach preserves size consistency, a property that isotropic rescaling along $\mathbf{p}$ violates. While we restricted our study of momentum rescaling to the TAB method, we believe the ideas presented here to be applicable to other stochastic Ehrenfest methods.

\balance

\section*{Conflicts of interest}
There are no conflicts to declare.
\section*{Data Availability}
The version of the TAB code that was used for calculations, along with code for the numerically exact dynamics for the 3-state model is available at https://github.com/blevine37/tab-model 
\section*{Acknowledgments}
This work was supported by U.S. Department of Energy, Office of Science, Office of Basic Energy Sciences, under Award No. DE-SC0021643 and by the Institute for Advanced Computational Science at Stony Brook University.

\bibliography{MyLibrary} 
\bibliographystyle{rsc} 

\end{document}